\documentclass[10pt,twocolumn,letterpaper]{article}

\usepackage{cvpr}              

\usepackage{graphicx}
\usepackage{multirow}
\usepackage{pifont}
\usepackage[normalem]{ulem}
\useunder{\uline}{\ul}{}
\usepackage{arydshln}
\definecolor{cvprblue}{rgb}{0.21,0.49,0.74}
\usepackage[pagebackref,breaklinks,colorlinks,allcolors=cvprblue]{hyperref}
\title{Towards Video to Piano Music Generation with Chain-of-Perform Support Benchmarks}
\author{
Chang Liu$^{1,2}$\thanks{Equal contribution.} \quad 
Haomin Zhang$^{1}$\footnotemark[1] \quad 
Shiyu Xia$^{1}$\footnotemark[1] \quad 
Zihao Chen$^{1}$ \quad 
Chaofan Ding$^{1}$ \\
Xin Yue$^{1}$ \quad 
Huizhe Chen$^{1}$ \quad 
Xinhan Di$^{1}$ \\
$^1$AI Lab, Giant Network \quad 
$^2$University of Trento \\
{\tt\small chang.liu-2@unitn.it, xiashiyu@ztgame.com, zhanghaomin@ztgame.com} \\
{\tt\small \{chenzihao, dingchaofan, v-yuexin, v-chenhuizhe, dixinhan\}@ztgame.com}
}

\begin{document}
\maketitle
\begin{abstract}
Generating high-quality piano audio from video requires precise synchronization between visual cues and musical output, ensuring accurate semantic and temporal alignment. However, existing evaluation datasets do not to fully capture the intricate synchronization required for piano music generation. A comprehensive benchmark is essential for two primary reasons: (1) Existing metrics fail to fully capture the complexity of video-to-piano music interactions, and (2) a dedicated benchmark dataset can provide valuable insights to accelerate progress in high-quality piano music generation. To address these challenges, we introduce the CoP Benchmark Dataset—a fully open-sourced, multi-modal benchmark designed specifically for video-guided piano music generation. The proposed chain-of-perform(CoP) benchmark dataset offers several compelling features: 1) Detailed multi-modal annotations: comprehensive labels facilitate precise semantic and temporal alignment between video content and piano audio, leveraging step-by-step (Chain-of-Perform) guidance. 2) Versatile evaluation framework: It enables rigorous evaluation of both general-purpose and specialized piano music generation tasks from videos. 3) Full open-sourcing: The complete dataset, including annotations and evaluation protocols, is publicly available at \url{https://github.com/acappemin/Video-to-Audio-and-Piano}, with continuous leaderboard updates to drive ongoing research in video-to-piano music generation.
\end{abstract}    
\section{Introduction}
Foley, the art of synthesizing ambient sounds and sound effects guided by videos, is fundamental to achieving high-quality audio that is both semantically aligned and temporally synchronized with the visual content \cite{luo2023difffoley,zhang2024foleycrafter,SpecVQGAN_Iashin_2021,wang2024frieren,cheng2024taming}. Existing foley and video-to-audio (V2A) models primarily focus on enhancing alignment through either specialized modules \cite{zhang2024foleycrafter,wang2023v2amapperlightweightsolutionvisiontoaudio,li2024tri} or unified model architectures \cite{cheng2024taming,chen2024video}. However, current video2audio \cite{cheng2024taming,chen2024video} and video2music(V2M) \cite{tian2025audiox} methods seldom are trained on audio-visual data reflecting sound domains—such as piano music, violin music  ~\cite{Lee2019ObservingPA, Koepke2020SightTS}—and rarely adopt structured, step-by-step guidance essential for high-fidelity synthesis~\cite{Su2020AudeoAG}. 

To overcome these limitations, we present a comprehensive multimodal benchmark dataset designed specifically for evaluating video-to-piano music generation. Leveraging the Chain-of-Perform (CoP) framework, our benchmark offers step-by-step annotations that capture both the semantic context and temporal alignment between videos and piano audio. This standardized evaluation framework is designed to assess the performance of existing V2A and V2M models while also driving further research into adaptive, high-quality audio synthesis.
\begin{figure*}[!htb]
    \centering
    \includegraphics[width=0.90\textwidth]{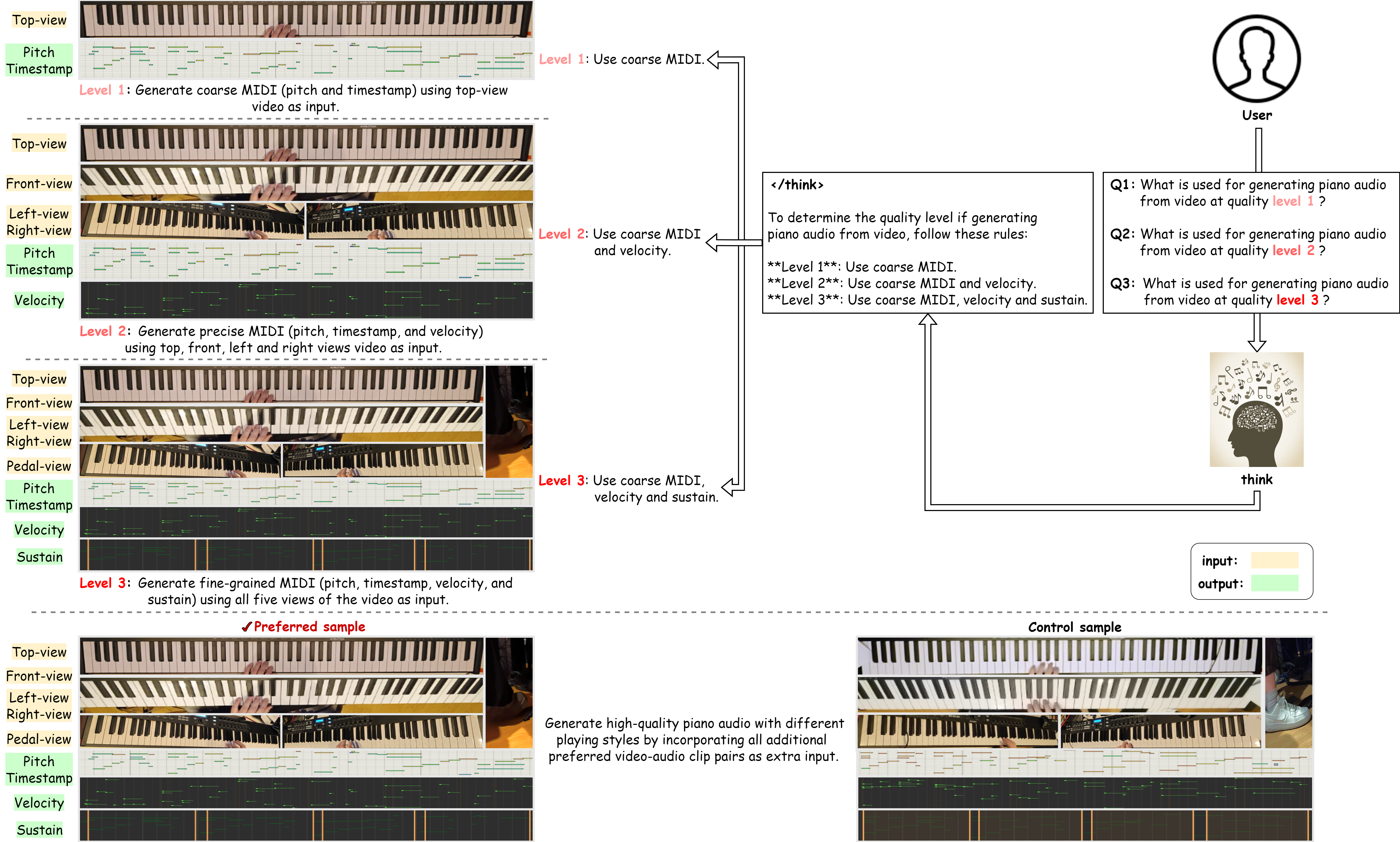}
    \caption{Video to Piano Music Generation with Chain-of-Perform(CoP) Dataset.}
    \label{views}
\end{figure*}

\section{Related Work}

\subsection{Video-to-Audio Synthesis and Multi-modal Reasoning}
Recent advancements in video-to-audio (V2A) synthesis have enabled high-quality Foley generation, progressing from autoregressive models~\cite{SpecVQGAN_Iashin_2021, viertola2024temporally, mei2024foleygen} to more efficient diffusion-based approaches~\cite{luo2023difffoley, wang2024frieren}. The incorporation of control signals such as timestamps and energy~\cite{zhang2024foleycrafter, li2024tri}, as well as multi-modal synchronization modules~\cite{chen2024video, cheng2024taming}, has improved alignment, though challenges remain in bridging audio-visual modality gaps in contexts. Meanwhile, visual piano transcription methods~\cite{Lee2019ObservingPA, Koepke2020SightTS, Su2020AudeoAG} show the potential of mapping video to MIDI through CNNs and GANs, but are often constrained by domain specificity and limited synchronization robustness. Complementing these efforts, multi-modal reasoning with Chain-of-Thought (CoT) models~\cite{xu2024llava, zhang2024improve, rafailov2024direct} provides structured integration across modalities, though their use in guiding V2M generation remains underexplored. These factors collectively drive the need for a unified benchmark that evaluates not only synthesis quality but also alignment in specialized scenarios, such as video-to-piano music generation.

\subsection{Evaluation of Video2Piano Music Generation}
Visual-to-audio models are typically evaluated with metrics like Fréchet Audio Distance (FAD), Fréchet Distance (FD), Inception Score (IS), and CLAP score ~\cite{chen2024video, cheng2024taming}, which assess fidelity, diversity, and semantic consistency. While these indicators are effective for general audio generation, they often fail to capture the alignment needed in context-sensitive tasks, such as video-to-piano music synthesis. Video-to-music (V2M) tasks require not only perceptual quality but also exact temporal and semantic alignment between visual content and musical structure. Existing methods often lack this degree of adaptivity. To address this, we combine standard V2A metrics with alignment-sensitive tasks specifically designed for musical performance, allowing for a more comprehensive evaluation of both the fidelity and expressiveness of generated piano music. 
\begin{table*}[t!]
\centering
\caption{Initial V2M Evaluation Metrics Towards Thinking}
\renewcommand{\arraystretch}{1.2}
\resizebox{\linewidth}{!}{
\begin{tabular}{ccccc|cccccc}
\toprule
\multicolumn{2}{c}{\multirow{2}{*}{\textbf{Model Name}}} 
& \multicolumn{3}{c|}{\textbf{Thinking Stage}} 
& \multicolumn{5}{c}{\textbf{Generation Stage}} \\
\cmidrule(lr){3-5} \cmidrule(lr){6-10}
\multicolumn{2}{c}{} 
& \textbf{Params} 
& \textbf{Format Score $\uparrow$} 
& \textbf{Outcome Score $\uparrow$} 
& \textbf{FD\textsubscript{PANNs} $\downarrow$} 
& \textbf{FD\textsuperscript{std}\textsubscript{PANNs} $\downarrow$} 
& \textbf{FD\textsubscript{PaSST} $\downarrow$} 
& \textbf{FD\textsuperscript{std}\textsubscript{PaSST} $\downarrow$} 
& \textbf{KL} \\
\midrule
\multicolumn{10}{c}{\textbf{Zero-shot}} \\
\midrule
\multicolumn{2}{c}{MMAudio-S-16kHz~\cite{cheng2024taming}} & 157M & - & - & 1.19 & 0.04 & 1.08 & 0.01 & 0.0246 \\
\multicolumn{2}{c}{MMAudio-S-44.1kHz~\cite{cheng2024taming}} & 157M & - & - & 1.14 & 0.05 & 1.09 & 0.03 & 0.0310 \\
\multicolumn{2}{c}{MMAudio-M-44.1kHz~\cite{cheng2024taming}} & 621M & - & - & 1.21 & 0.09 & 1.13 & 0.04 & 0.0317 \\
\multicolumn{2}{c}{MMAudio-L-44.1kHz~\cite{cheng2024taming}} & 1.03B & - & - & 1.13 & 0.04 & 1.09 & 0.03 & 0.0327 \\
\multicolumn{2}{c}{Yingsound~\cite{chen2024yingsound}} & - & - & - & 1.26 & 0.09 & 1.21 & 0.04 & 0.0259 \\
\multicolumn{2}{c}{FolyCrafter~\cite{zhang2024foleycrafter}} & - & - & - & 1.14 & 0.04 & 1.10 & 0.03 & 0.0276 \\
\midrule
\multicolumn{10}{c}{\textbf{Finetuned}} \\
\midrule
\multicolumn{2}{c}{MMAudio-S-16kHz~\cite{cheng2024taming}} & 157M & 100 & 100 & 1.19 & 0.04 & 1.08 & 0.01 & 0.0246 \\
\multicolumn{2}{c}{MMAudio-S-44.1kHz~\cite{cheng2024taming}} & 157M & 100 & 100 & 1.26 & 0.09 & 1.21 & 0.04 & 0.0259 \\
\multicolumn{2}{c}{MMAudio-M-44.1kHz~\cite{cheng2024taming}} & 621M & 100 & 100 & 1.19 & 0.03 & 1.17 & 0.03 & 0.0246 \\
\multicolumn{2}{c}{MMAudio-L-44.1kHz~\cite{cheng2024taming}} & 1.03B & 100 & 100 & 1.14 & 0.04 & 1.10 & 0.03 & 0.0276 \\
\multicolumn{2}{c}{Yingsound~\cite{chen2024yingsound}} & - & 100 & 100 & 1.115 & 0.049 & 1.056 & 0.018 & 0.0249 \\
\multicolumn{2}{c}{FolyCrafter~\cite{zhang2024foleycrafter}} & - & 100 & 100 & 1.274 & 0.03 & 1.092 & 0.01 & 0.0308 \\
\bottomrule
\end{tabular}
}
\label{tab:musicgenerationmetrics}
\end{table*}

\section{Benchmark Suite}

\subsection{Towards Multi-Modal CoP Dataset Suite}
We have constructed a 10-hour multi-modal video-to-piano CoT-like (CoP) dataset for generating high-quality piano music from videos. The primary constraint for data collection was a five-view piano performance with a fully visible keyboard and practice pedal. We employed two skilled pianists with distinct performance styles to record this dataset. Inspired by Chain-of-Thought (CoT) and CoT-like guidance~\cite{wei2022chain}, we further developed a step-by-step pipeline that separates the reasoning and generation processes into two main stages, as illustrated in Fig.~\ref{views}.

\noindent
\textbf{Thinking Stage.} In this stage, a fine-tuned large language model interprets the user’s query—such as specifying the desired piano music quality or style—and determines the required components (e.g., MIDI pitch, velocity, sustain) along with the relevant camera views needed for each generation level. This chain-of-thought-like(CoP) guidance serves as an explicit blueprint, indicating precisely which visual inputs or reference clips to use.

\noindent
\textbf{Generation Stage.} Guided by the structured instructions from the Thinking Stage, our benchmark supports the generation of piano music across four progressively refined levels:

\begin{enumerate}
    \item \textbf{Level 1:} Generate coarse MIDI (pitch and timestamp) using only the top-view video.
    \item \textbf{Level 2:} Generate precise MIDI (pitch, timestamp, and velocity) using top, left, right, and front views.
    \item \textbf{Level 3:} Generate fine-grained MIDI (pitch, timestamp, velocity, and sustain) by incorporating all five views, including the pedal view.
    \item \textbf{Level 4:} Produce high-quality piano audio with different playing styles, leveraging additional preferred video-audio clip pairs as extra inputs.
\end{enumerate}

Our benchmark allows for more transparent, controllable, and high-fidelity video-to-piano music synthesis. As shown in Fig.~\ref{views}, the expert-provided annotations in each step guide the model towards high-standard performance, ensuring robust alignment between visual cues and musical output.

\subsection{Evaluation Metrics Suite}
We employ two sets of metrics to comprehensively assess our system towards both the reasoning (Thinking Stage) and music generation (Generation Stage) perspectives.

\noindent
\textbf{Thinking Metrics.} 
During the Thinking Stage, we evaluate the performance of our trained large language reasoning model (LLM) as the following:
\begin{itemize}
    \item \textbf{Form Accuracy:} Evaluates the structural correctness of the model’s output (e.g., whether required components or steps are properly listed).
    \item \textbf{Outcome Accuracy:} Checks whether the final reasoning result aligns with the ground truth or expert annotations.
\end{itemize}

\noindent
\textbf{Music Generation Metrics.}
For evaluating semantic alignment, temporal alignment, and audio quality on the VGGSound and AudioCaps test sets, we adopt standard metrics including Inception Score (IS)~\cite{salimans2016improved}, CLIP score, Fréchet Distance (FD)~\cite{heusel2017gans}, Fréchet Audio Distance (FAD), AV-align (AV)~\cite{yariv2024diverse}, KL Divergence-softmax (KL-softmax)~\cite{SpecVQGAN_Iashin_2021}, and CLAP score~\cite{wu2023large}. These indicators capture the overall fidelity, diversity, and semantic relevance of the generated audio. In the context of piano music, we further measure the precision, recall, accuracy of MIDI. Multiple evaluators provide Mean Opinion Scores (MOS) for these assessments, offering insights into the perceived smoothness and artistic appeal of the generated music independent of any reference.

\section{Experiments}
We conduct initial experiments on a variety of state-of-the-art models, including V2A models, large language reasoning models. We will continue our experiments on a variety of the state-of-the-art V2M models. As shown in Table \ref{tab:musicgenerationmetrics}, the format score and outcome score are $100\%$ for all state-of-the-art reasoning models(open-source), including DeepSeek-R1 models(From R1-670B to Qwen-Distill-1.5B)\cite{guo2025deepseek}, Qwen-QwQ-32B\cite{qwen2025qwen25technicalreport}. Both the format score and the outcome score are $100\%$ after training, while the format score and the outcome score is very low for zero-shot inference. The evaluation generation metrics are demonstrated in the Table \ref{tab:musicgenerationmetrics}, although the performance increase a bit after trained on a variety of state-of-the-art V2A models, the quality of the generated piano music is still not satisfactory. Besides, the evaluation performance on MIDI and MOS are very low.   

\section{Discussion}
We propose the towards video to piano music generation with chain-of-perform support benchmarks, and initial experiments are conducted on a variety of the state-of-the-art models. We will further conduct experiments on the state-of-the-art V2M models and develop new models for high-quality music generation.  


{
    \small
    \bibliographystyle{ieeenat_fullname}
    \bibliography{main}
}


\end{document}